\documentclass[%
reprint, 
superscriptaddress,
groupedaddress,
nofootinbib,
amsmath,amssymb,
aps,
]{revtex4-1}

\usepackage[normalem]{ulem}
\usepackage{amsmath,amssymb}
\usepackage{graphicx}
\usepackage{dcolumn}
\usepackage{bm}
\usepackage[colorlinks=true, linkcolor = red, citecolor = blue]{hyperref}
\usepackage[T1]{fontenc}
\usepackage{booktabs}

\begin{document}


\title{Nonlinear cosmological structure with ultralight bosons via modified gravity}

\author{Stefany G. Medell\'in-Gonz\'alez} 
 \email{sg.medellingonzalez@ugto.mx}
\affiliation{%
Departamento de F\'isica, DCI, Campus Le\'on, Universidad de
Guanajuato, 37150, Le\'on, Guanajuato, M\'exico.}

\author{Alma X. Gonz\'alez-Morales}%
 \email{alma.gonzalez@fisica.ugto.mx}
 \affiliation{Consejo Nacional de Ciencia y Tecnolog\'ia,
Av. Insurgentes Sur 1582. Colonia Cr\'edito Constructor, Del. Benito   Ju\'arez C.P. 03940, M\'exico D.F. M\'exico}
\affiliation{%
Departamento de F\'isica, DCI, Campus Le\'on, Universidad de
Guanajuato, 37150, Le\'on, Guanajuato, M\'exico.}

\author{L. Arturo Ure\~{n}a-L\'{o}pez} 
 \email{lurena@ugto.mx}
\affiliation{%
Departamento de F\'isica, DCI, Campus Le\'on, Universidad de
Guanajuato, 37150, Le\'on, Guanajuato, M\'exico.}

\date{\today}

\begin{abstract}
Ultralight bosons as dark matter has become a model of major interest in cosmology, due to the possible imprint of a distinct signature in the cosmic structure both at the linear and nonlinear scales. In this work we show that the equations of motion for density perturbations for this kind of models can be written in terms of a modified gravitational potential. Taking advantage of this parallelism, we use the \texttt{MG-PICOLA} code originally developed for modified gravity models to evolve the density field of axion models with and without self-interaction. Our results indicate that the quantum potential adds extra suppression of power at the nonlinear level, and it is even capable of smoothing any bumpy features initially present in the mass power spectrum.

\end{abstract}

\pacs{Valid PACS appear here}
\maketitle

\section{Introduction \label{sec:intro}}
The existence of dark matter (DM) in the Universe is one of the most challenging puzzles of modern cosmology\cite{Baudis2018} and more so because it has been difficult to determine its intrinsic properties beyond the assumption of a pressureless fluid made of massive particles. It is then no surprise that one alternative model, to the so-called cold dark matter (CDM) one,  has captured the attention of the community in recent years\cite{Niemeyer:2019aqm,Urena-Lopez:2019kud,Hui2017,Marsh2016,Magana2012,*Magana2012a}, mostly because of the intriguing possibility that the quantum properties of its constituent particles could have been imprinted in the features of galaxies and large scale structure in the cosmos. The model has been known by different names in the literature, and then we will refer to it indistinctly as scalar field dark matter (SFDM)\cite{Matos2000,*Matos2001,*Matos2000a}, fuzzy dark matter (FDM)\cite{Hu2000}, or ultralight axions\cite{Marsh2016}.

As a direct detection of this kind of particle seems to be beyond the reach of our present technological capabilities\cite{Manley:2019vxy,Emken2018,Aprile2017,Cui2017}, we must rely on cosmological (i.e., gravitational) observations as the best option to test its existence (e.g.,~\cite{Grin:2019mub,Poulin:2018dzj}). So far, the cosmological evolution is well understood up to the level of linear perturbations, with even dedicated versions of the leading Boltzmann codes that allow a precise comparison of the model with a full suite of cosmological observations\cite{cedeo2020ultralight,Cedeno2017,*Urena-Lopez2016,Hlozek:2017zzf,*Hlozek:2016lzm,*Hlozek:2014lca,Cookmeyer:2019rna}. 

However, the nonlinear formation of structure still is a difficult task, and it has been until recently that full simulations seem to be at hand for a proper study of the formation of structure under axion models\cite{schwabe2020axionyx,Mocz:2019uyd,*Mocz:2019emo,*Lancaster:2019mde,*Mocz2018,*Mocz:2017wlg,Nori2018,*Nori:2018pka,Schive2014,Woo2009}. A summary of simulations produced by different groups is shown in Table~\ref{tab:sims}, one can see that the typical size of the simulations box is of the order of tens of $\mathrm{Mpc}/h$ and boson mass of around $10^{-22} \mathrm{eV}/c^2$. Although these kind of simulations will allow the systematic testing of the model in the near future, it is necessary to find options for the generation of rapid cosmological simulations much in the same way as it has been done for CDM and modified gravity (MG) models\cite{Vogelsberger:2019ynw,Valogiannis2017,Winther2017,Aviles2017}.

\begin{table*}
    \centering
    \caption{\label{tab:sims} Summary of some of the existing FDM cosmological simulations. From left to right, the columns refer to: the reference in the literature; the mass of the boson particle $m_{a}$; the numerical method used for the simulation, where SP stands for the Schrodinger-Poisson system and SPH for smooth particle hydrodynamics; the box size of the simulation and the number of particles $N_{\rm part}$. 
    }  
    \small
    \begin{tabular}{*{5}{p{.15\linewidth}}}
      \toprule
    Ref. & $m_a \, (\mathrm{eV}/c^2) $&  Method & Box size $(\mathrm{Mpc}/h)$ & $N_{\rm part}$ \\\midrule
    \midrule         
    \textbf{May \& Springel (2021) }\cite{may2021structure} & $3.\times 10^{-23}$, $7\times 10^{-23}$ & SP & $10$,$5$ & $8640^{3}$, $4320^{3}$, $3072^{3}$, $2048^{3}$  \\\midrule
    \textbf{Schwabe et al (2020)} \cite{schwabe2020axionyx} & $10^{-22}$& SP &2$h$&$1024^{3}$ \\\midrule
    \textbf{Li et al(2019)}  \cite{Li_2019} &  $10^{-22}$, $10^{-23}$& SP vs SPH &$20$, $10$, $3$ & $1024^3$\\\midrule
    \textbf{Mocz et al. (2018)} \cite{Mocz2018}& $10^{-22}, 10^{-21}$&SP vs SPH & $0.250$ &$1024^{3}$  \\ \midrule
    \textbf{Nori et al. (2018) }\cite{Nori2018} &$10^{-22}$& SPH &$10, 15$ & $256^3$, $512^3$ \\\midrule
    \textbf{Zhang et al(2018) } \cite{Zhang2017} &$10^{-22}$ & SPH &$0.400$ & $10^6$ \\\midrule
    \textbf{Woo \& Chiueh (2009) } \cite{Woo2009} &$10^{-22}$& SP &1& $1024^3$\\\midrule
    \bottomrule
    \end{tabular}
  \end{table*}

In this paper we explore such possibility for axion models and argue that, given the scale-dependent growth of density perturbations in these types of models we can use the  recently developed  Comoving Lagrangian Acceleration (COLA) method\cite{Tassev2013}, in the form originally developed for MG models \texttt{MG-PICOLA}\footnote{https://github.com/HAWinther/MG-PICOLA-PUBLIC}\cite{Winther2017},  (see also~\cite{howlett2015lpicola,tassev2015scola}). We will do this for the standard FDM model, with its distinctive cutoff in the linear mass power spectrum (MPS), and also for the so-called extreme axion model\cite{Cedeno2017,Zhang2017b,Zhang2017c} that features an excess of power, when compared to CDM, in the growth of linear density perturbations. A key intermediate step in the calculations will be the transformation of the axion field equations into their hydrodynamic counterpart in terms of the well-known Madelung transformation.

A summary of the paper is as follows. In Sec.~\ref{sec:math-back} we present the main equations of motion for axion models in the nonrelativistic limit, and correspondingly their final form in terms of hydrodynamic quantities. We will also explain the expected growth of linear density perturbations and the corresponding appearance of two characteristic length scales (and their corresponding wave numbers) related to the physical parameters in the model. From here we write the growth factors that describe the evolution of density perturbations up to the second order in Lagrangian dynamics, and show how their equations of motion are modified to include the so-called quantum pressure that arises because of the wave properties of axion dark matter. These results will be implemented in \texttt{MG-PICOLA}\cite{Winther2017} for scale-dependent models (see also\cite{Bose2016,Aviles2017}).

In Sec.~\ref{sec:cosmo-sims} we describe the cosmological simulations that are obtained from the theoretical framework of Sec.~\ref{sec:math-back}, and for different cosmological setups in terms of initial conditions and evolution equations, so that we can determine the main contributions to the nonlinear evolution of density perturbations of the axion models. The initial conditions consider the usual FDM suppression of power at small scales, but we also include the possibility for the extreme case for which there is an excess of power as compared to the CDM case, just before the abrupt suppression. Finally, in Sec.~\ref{sec:conclusions} we discuss the main results and future perspectives of this work.

\section{Mathematical background \label{sec:math-back}}
The equations of motion of ultralight axions in the nonrelativistic approximation have been obtained before in Refs.~\cite{Marsh2015,Suarez2017,Suarez2017a,Fan2016,Urena-Lopez2014,Guzman2004,Seidel1990}, here we give a brief description of their derivation. Let us start with the following action,
\begin{equation}
S\!=\! \int\! \sqrt{-g} \, d^4 x\! \left[ \frac{R}{2 \kappa^2}\! + \partial_\mu \phi \partial^\mu \phi^\ast + \tilde{m}^2_a |\phi|^2 + \tilde{\lambda} |\phi|^4  \right] ,
\label{eq:action}
\end{equation}
where $g = \mathrm{det}(g_{\mu \nu})$, $\kappa^2 = 8\pi G/c^4$, $G$ is Newton's gravitational constant, $c$ is the speed of light, $R$ is the Ricci scalar and $\phi$ is a complex scalar field endowed with a potential that contains quadratic and quartic interactions terms.\footnote{Although there are some differences between real and complex scalar fields at the relativistic level, their nonrelativistic dynamics, which is the regime of physical interest here, is driven by the same set of equations. Thus, all results presented here apply for both the real and complex cases.} 

The parameters in the potential are explicitly given by $\tilde{m}_a = m_a c/\hbar$ and $\tilde{\lambda} = \lambda/(\hbar c)$, where $\hbar$ is the Planck's constant, $m_a$ is the boson mass and $\lambda$ is a dimensionless parameter. Notice that the fundamental constants will be shown explicitly, and that the potential parameters become the bare ones if we choose to use natural units $c= \hbar=1$, i.e $\tilde{m}_a = m_a$ and $\tilde{\lambda} = \lambda$.\footnote{Under our convention the units of the different physical quantities are as follows: $[\phi] = [\psi] = (\mathrm{energy/length})^{1/2}$, $[\tilde{m}_a] = \mathrm{length}^{-1}$ and $[\tilde{\lambda}]= (\mathrm{energy \cdot length})^{-1}$. With this Planck's constant $\hbar$ will not appear explicitly in Eqs.~\eqref{eq:fluid-qbep} below. In natural units we find that: $[\phi] = \mathrm{energy}$, $[\tilde{m}_a] = \mathrm{energy}$ and $\tilde{\lambda}$ becomes dimensionless.} Notice that the Compton length of the boson particle is just $L_C = 1/\tilde{m}_a$, and correspondingly the Compton time is defined as $T_C = L_C/c$, which is the one used in the field transformation that leads to Eqs.~\eqref{eq:gpp-equations}.

For the nonrelativistic limit (for details see~\cite{Fan2016,Urena-Lopez2014,Chavanis2011,Guzman2004,Seidel1990,Ruffini1969}), we now apply the field transformation $\phi(t,\mathbf{r}) = e^{i \tilde{m}_a c t} \psi(t,\mathbf{r})$, where the new field $\psi$ will obey the conditions that the order of magnitude of the field derivatives are $\partial_t \psi \sim \partial^2_r \psi = \mathcal{O}(\epsilon^2)$, with $|\epsilon| \ll 1$. Additionally, we consider a perturbed FRW metric in the form $ds^2 = -(1+2\Psi/c^2) dt^2 + a^2(t) (1-2\Psi/c^2) d\mathbf{x}^2$, where $\Psi$ is the Newtonian gravitational potential and $a=a(t)$ is the scale factor of the Universe. Thus, the equations of motion derived from the action~\eqref{eq:action} can be written in the form of the well-known Gross-Pitaevskii-Poisson (GPP) system,
\begin{subequations}
\label{eq:gpp-equations}
{\small
\begin{eqnarray}
i \hbar \frac{\partial_t (a^{3/2} \psi)}{a^{3/2}}\! =\! - \frac{\hbar c}{2\tilde{m}_a a^2} \nabla^2 \psi +\! \left( \frac{\Psi}{c^2} +\! \frac{\tilde{\lambda}}{2\tilde{m}^2_a} |\psi|^2 \right)\! \tilde{m}_a \psi\, ,  \label{eq:gpp-equations-a}& \\
\nabla^2 \Psi = \frac{4\pi G}{c^2} a^2 \tilde{m}^2_a |\psi|^2 \left( 1 + \frac{\tilde{\lambda}}{2\tilde{m}^2_a} |\psi|^2 \right) - \frac{3}{2} a^2 H^2 \, , \label{eq:gpp-equations-b}&
\end{eqnarray}
}
\end{subequations}
where $H$ is the Hubble parameter. Eqs.~\eqref{eq:gpp-equations} are complemented by the Friedmann constraint $H^2 = (8\pi G/c^2) \rho_b$, being $\rho_b$ the background (homogeneous) density.

One note is in turn regarding the quadratic terms in Eqs.~\eqref{eq:gpp-equations}, which come from the quartic field term in the action~\eqref{eq:action}. As shown in a variety of previous works\cite{Li2013,Lin2018}, the quartic term changes the early-time behavior of the scalar field, when the latter cannot be considered pressureless. Hence, one comes to the general conclusion that the quadratic term in Eqs.~\eqref{eq:gpp-equations-b} should become small before the expected time of radiation-matter equality as otherwise the field would be unable to behave as a proper DM component at late times. In other words, once the scalar field behaves as pressureless matter, we find that $(\tilde{\lambda}/2\tilde{m}^2_a) |\psi|^2 \ll 1$ for the time window of structure formation (which is the time of main interest here), and then such term could be dropped from Eq.~\eqref{eq:gpp-equations-b}, but we will come back to this point later. 

We now apply the Madelung transformation in the form $\tilde{m}_a \psi = c \sqrt{\rho(t,\mathbf{x})} e^{i S(t,\mathbf{x})/\hbar}$, where $\rho (t,\mathbf{x})$ is the density field and $\mathbf{u} = \nabla S/m_a$ is the velocity field, which is then irrotational by definition, $\nabla \times \mathbf{u} = 0$. It must be noted that in doing the Madelung transformation we are assuming that the nonrelativistic matter density is strictly given by $\rho = \tilde{m}^2_a |\psi|^2/c^2$. The GPP system~\eqref{eq:gpp-equations} then becomes the quantum barotropic Euler-Poisson (QBEP) one\cite{matos2009scalar,Chavanis2011,*Chavanis2012,*Suarez2011,*Suarez2017,rodrguezmeza2017materia,Nori2018,Mocz2018}, which we write in the form
\begin{subequations}
\label{eq:fluid-qbep}
\begin{eqnarray}
\frac{\partial \rho}{\partial t} + 3H\rho + \frac{1}{a^2} \nabla \cdot (\rho \mathbf{u}) = 0 \, ,& \label{eq:fluid-qbep-a} \\
\frac{\partial \mathbf{u}}{\partial t} + (\mathbf{u} \cdot \nabla) \mathbf{u} =  - \nabla \tilde{\Psi} \, ,& \label{eq:fluid-qbep-b} \\
\nabla^2 \tilde{\Psi} = 4\pi G a^2 \rho + \frac{\tilde{\lambda} c^4}{2\tilde{m}^4_a} \nabla^2 \rho - \frac{c^2}{2\tilde{m}^2_a} \frac{\nabla^2}{a^2} \left( \frac{\nabla^2 \sqrt{\rho}}{\sqrt{\rho}} \right) \, . & \label{eq:fluid-qbep-c}
\end{eqnarray}
\end{subequations}
To serve our purposes, we have written the force on the right-hand side (rhs) of Eq.~\eqref{eq:fluid-qbep-b} in terms of the gradient of an effective potential $\tilde{\Psi}$, whose corresponding Poisson equation~\eqref{eq:fluid-qbep-c} is sourced by a combination of linear and nonlinear functions of the matter density $\rho$. The first term on the rhs is the standard Newtonian source term, the second one comes from the quartic self-interaction in Eq.~\eqref{eq:gpp-equations-a}, and the last one is called the quantum potential. 

Notice that in a more standard approach, the last two terms on the rhs of Eq.~\eqref{eq:fluid-qbep-c} are considered as extra, internal forces, apart from the gravitational one, acting on the fluid elements; that is, as two new sources of acceleration~\cite{Nori:2018pka,Zhang2017,Mocz2015}. The CDM case is clearly obtained when those two sources are absent. In the rearrangement that we have made, the same picture is preserved with a twist: the evolution of individual particles under their own self-gravity, where the latter now considers two non-Newtonian sources of gravity. This will leave the system~\eqref{eq:fluid-qbep}, and in consequence the full boson system, in suitable form for its evolution by means of $N$-body techniques, see also~\cite{Vogelsberger:2019ynw,Chac_n_2020,Nori:2018pka,Li_2019} and references therein for detailed comparisons with solutions from the GPP system.

To understand the underlying properties of the new self-gravitating system, it is convenient to write Eqs.~\eqref{eq:fluid-qbep} in terms of the density contrast $\delta (t,\mathbf{x}) = \rho (t,\mathbf{x})/\rho_b -1$, whereas for the velocity we take $\mathbf{u} = H \mathbf{r} + \mathbf{v}$, with $\mathbf{v}$ the peculiar velocity. As is well known from the theory of linear density perturbations, corresponding to $|\delta|  \ll 1$  and $|\mathbf{v}| \ll 1$, Eqs.~\eqref{eq:fluid-qbep} can be combined into a single one for the density contrast in the form 
\begin{eqnarray}
\ddot{\delta} = 4 \pi G \rho_b a^4 \delta + \frac{\tilde{\lambda} c^4 \rho_b a^2}{\tilde{m}^4_a} \nabla^2 \delta - \frac{c^2}{4 \tilde{m}^2_a} \nabla^2 \left( \nabla^2 \delta \right)  \, , \label{eq:poissons-a}
\end{eqnarray}
where a dot means derivative with respect to the so-called superconformal time $\tau$ defined through the relation $a^2 d\tau =dt$.

Thus, the equation of motion~\eqref{eq:poissons-a} at linear order for a given Fourier mode of the density contrast $\delta_k$, with corresponding wave number $k$, reads
\begin{subequations}
\label{eq:qbep-mg}
\begin{equation}
    \ddot{\delta}_k = 4 \pi G \rho_b a^4 \mu(a,k,\tilde{\lambda}) \delta_k \, , \label{eq:qbep-mgcola-a}
\end{equation}
where \cite{refId0, Fan2016}
\begin{equation}
    \mu(a,k,\tilde{\lambda}) = 1 - k^2/k^2_{J1} - k^4/k^4_{J0} \, . \label{eq:qbep-mgcola}
\end{equation}
\end{subequations}
In the foregoing equation, we have implicitly defined the Jeans wave number of FDM as: $k^4_{J0} = (16 \pi G/c^2) \rho_b a^4 \tilde{m}^2_a$. Also, we have defined a second characteristic wave number as: $k^2_{J1} = (4 \pi G/c^4) a^2 \tilde{m}^4_a /\tilde{\lambda}$. For reference, the CDM case corresponds to $\mu(a,k,\tilde{\lambda})=1$.

The first Jeans wave number can be written as~\cite{Hu2000,Urena-Lopez2016,Marsh2016}\footnote{ Using the Friedmann equation $H^2 = (8\pi G/3) \rho_b$, where $H$ is the Hubble parameter, Eq.~\eqref{eq:jeans0} can also be written as $k_{J0} = 6^{1/4} a \sqrt{H \tilde{m}_a/c}$, which shows that $k^{-1}_{J0}$ is just the geometrical mean of the Hubble distance and the Compton length of the boson particle~\cite{Urena-Lopez2016}.}, 
\begin{equation}
    k_{J0}\! = 66.5 a^{1/4}\! \left( \frac{m_a c^2}{10^{-22} \mathrm{eV}} \right)^{1/2}\! \left( \frac{\Omega_{a0} h^2}{0.12} \right)^{1/4}  \mathrm{Mpc}^{-1} \, . \label{eq:jeans0}
\end{equation}
The influence of $k_{J0}$ in the evolution of density perturbations changes with time, and some level of suppression of the perturbations, in comparison with CDM, is expected for scales in the range $k \gtrsim 10 \, h \mathrm{Mpc}^{-1}$ ever since the time of radiation-matter equality. 

Likewise, the effects from the second instability scale $k_{J1}$ on the growth of density perturbations are only present at early times. One must recall that the second term on the rhs of Eq.~\eqref{eq:poissons-a} is related to the quartic term in the Schrodinger equation~\eqref{eq:gpp-equations-a}, and then we also find $(\tilde{\lambda}/2\tilde{m}^2_a) |\psi|^2 = (\tilde{\lambda} c^2/2 \tilde{m}^4_a) \rho_b \ll 1$ after the time of radiation-matter equality. The main conclusion is then that any effects from the second Jeans wave number $k_{J1}$ can be studied in terms of the fluid equations of the FDM case ($\tilde{\lambda}=0$), as long as the initial conditions already take into account the effects of the SF self-interaction on the linear MPS (see for instance Sec.~\ref{sec:cosmo-sims} below). Because of this, we will hereafter consider $\tilde{\lambda} =0$ in all expressions regarding the evolution of density perturbations.

Our next step is to study the nonlinear stage of structure formation, and for that, we resort to the scale-dependent COLA method of Ref.~\cite{Winther2017}. We first assume that the equation of motion of the particles in an $N$-body simulation is $ \ddot{\mathbf{x}} = - \nabla \tilde{\Psi}$, where $\tilde{\Psi}$ is the potential in Eqs.~\eqref{eq:fluid-qbep}. From this we obtain a scale-dependent, first order, growth factor $D_1(\tau,k)$ that obeys the equation
\begin{subequations}
\label{eq:growths}
\begin{equation}
   \ddot{D}_1 - 4 \pi G \rho_b a^4 \mu(a,k) D_1 = 0 \, . \label{eq:growths-a}
\end{equation}
A second order growth factor $D_2(\tau, k)$ is also obtained by simply relying on the expressions developed in\cite{Winther2017} [see their Eq.~(4.10)] for modified gravity models, namely,
\begin{widetext}
\begin{equation}
    \ddot{D}_2 - 4 \pi G \rho_b a^4 \mu(a,k) D_2 = - D^2_1 \left[ 4 \pi G \rho_b a^4 \mu (a,k) + 2 a^4 H^2 \gamma_2 (\mathbf{k},\mathbf{k}_1,\mathbf{k}_2,a ) \right] \, , \label{eq:growths-b}
\end{equation}
where
\begin{equation}
    \gamma_2(\mathbf{k},\mathbf{k}_1,\mathbf{k}_2,a) = \gamma^E_2 (\mathbf{k},\mathbf{k}_1,\mathbf{k}_2,a) + \frac{3}{4} \frac{\mathbf{k}_1 \cdot \mathbf{k}_2}{k^2_1 k^2_2} \Omega_b(a) \left[ \left( \mu(a,k) - \mu(a,k_1) \right) k^2_1 + \left( \mu(a,k) - \mu(a,k_2) \right) k^2_2 \right] \, , \label{eq:growths-c}
\end{equation}
\end{widetext}
\end{subequations}
Here, $\mathbf{k}_1$ and $\mathbf{k}_2$ are wave vectors in Fourier space that must comply with the triangle constraint $\mathbf{k} = \mathbf{k}_1 + \mathbf{k}_2$.

To find $\gamma^E_2$ we require an expansion up to second order of $\nabla^2 \tilde{\Psi}$ in Fourier space. In configuration space, we find for the third term on the rhs of Eq.~\eqref{eq:fluid-qbep-c},
\begin{widetext}
\begin{eqnarray}
    \nabla^2 \left( \frac{\nabla^2 \sqrt{1+\delta}}{\sqrt{1+\delta}} \right) &=& \frac{1}{2} \nabla^2(\nabla^2 \delta) - \frac{1}{8} \nabla^2 (\nabla^2 \delta^2) - \frac{1}{4} (\nabla^2 \delta)^2 - \frac{1}{2} \nabla \delta \cdot \nabla (\nabla^2 \delta) - \frac{1}{4} \delta \nabla^2 (\nabla^2 \delta) + \mathcal{O}(\delta^3) \, . \label{eq:second-order}
\end{eqnarray}
\end{widetext}

Following the notation of\cite{Winther2017,Bose2016}, the second order kernel $\gamma^E_2$ obtained from the Fourier transform of Eq.~\eqref{eq:second-order} is 
\begin{equation}
    \gamma^E_2 = \frac{c^2}{8a^4 H^2 \tilde{m}^2_a} (k^2_1 + k^2_2 + \mathbf{k}_1 \cdot \mathbf{k}_2) (\mathbf{k}_1 + \mathbf{k}_2)^2 \, . \label{eq:second-kernel}
\end{equation}

\section{Cosmological simulations \label{sec:cosmo-sims}}
To study the nonlinear evolution of density perturbations, we then use the code \texttt{MG-PICOLA}\cite{Winther2017} supplied with the MG functions $\mu(a,k)$ and $\gamma^E_2(a,k)$ described above. But before we describe the main results, there are a couple of pertinent notes about the numerical calculations in the code.

The first one is related to the solution of the Fourier integrals. To enhance the speed of the calculations in Fourier space, \texttt{MG-PICOLA} takes by default the orthogonal triangle configuration $\mathbf{k}_1 \cdot \mathbf{k}_2 =0$ and $k_1 = k_2 = k/\sqrt{2}$~\cite{Winther2017}. This is the same choice considered in our simulations, under which we find from Eq.~\eqref{eq:growths-c} that $\gamma_2 = \gamma^E_2 = c^2 k^4/(8 a^4 H^2 \tilde{m}^2_a)$. The latter implies that the second term on the rhs of Eq.~\eqref{eq:growths-b} contributes with the same factor as that of the Jeans wave number in the first term, but with the opposite sign. The overall result is then that the equation of motion of the second growth factor~\eqref{eq:growths-b} simplifies to
\begin{equation}
     \ddot{D}_2 - 4 \pi G \rho_b a^4 \mu(a,k) D_2 = -  4\pi G \rho_b a^4 D^2_1 \, . \label{eq:growths-d}
\end{equation}

Thus, the only scale-dependent influence on the evolution of $D_2$ is provided by the term $\mu(a,k)$. It must be stressed out that this result arises directly from both the nature of the SFDM model and the triangle approximation to increase the speed of \texttt{MG-PICOLA}. Other approximations, like the so-called equilateral ($k_1 = k_2 =k$ and $\mathbf{k}_1 \cdot \mathbf{k}_2 =k^2/\sqrt{2}$) or squeezed ($k_1 = k$, $k_2 \simeq 0$ and $\mathbf{k}_1 \cdot \mathbf{k}_2 \simeq 0$) ones, may allow for more scale-dependent effects in the evolution of $D_2$. However, as argued in~\cite{Winther2017}, it is expected that the orthogonal approximation already takes into account most of the weight arising from the Fourier integrals.

The second note is about the initial conditions for the growth factors. Once supplied with a linear MPS of the desired model at the initial redshift $z_i$, the code evolves the growth factors $D_1$ and $D_2$, by means of Eqs.~\eqref{eq:growths}, up to the present time using the initial conditions $D_1(z_i,k) = 1$, $\dot{D}_1(z_i,k) = \mathcal{H}_i$, $D_2(z_i,k)=-3/7$ and $\dot{D}_2(z_i,k) = -(6/7) \mathcal{H}_i$, where $\mathcal{H}= \dot{a}/a$. The resultant growth factors are then normalized so that their values at $z=0$ is unity, that is, $D_1(0,k) =1 = D_2(0,k)$. This is a convenient procedure for the CDM case, for which the growth factors are scale independent, and then the normalization at $z=0$ translates into the same initial conditions at $z_i$ for all scales.

For models with scale-dependent growth factors, as our model of interest, the normalization to unity at $z=0$ means that the initial conditions at $z_i$ are not the same for all scales, but that the initial amplitudes of $D_1(z_i,k)$ and $D_2(z_i,k)$ would be artificially deformed and may not be the correct ones corresponding to the scale-dependent model under consideration. To minimize the deformation of the initial conditions, we have instead normalize the growth factors with respect to their values at large scales, ie $D^{new}_1(z,k) = D_1(z,k)/D_1(0,k \to 0)$ and $D^{new}_2(z,k) = D_2(z,k)/D_2(0,k \to 0)$. As our model has the same perturbations at large scales as CDM, the normalization factors are also the same obtained from a CDM simulation, that is, $D_1(0,k \to 0) = D^{CDM}_1(0)$ and $D_2(0,k \to 0) = D^{CDM}_2(0)$. In this form, the initial conditions will be the same for all scales, as we readily find that $D^{new}_1(z_i,k) = 1/D_1(0,k \to 0)$ and $D^{new}_2(z_i,k) = -3/(7D_2(0,k \to 0))$, and any scale-dependent power suppression is left encoded intact in the initial MPS. 

 Following the labeling in\cite{Zhang2017,Nori2018} for cosmological simulations, we studied the following cases: CIC (Cold Initial Conditions)\footnote{Not to confuse with the standard notation CIC in $N$-body algorithms that stands for Cloud In Cell.}, that corresponds to a typical CDM simulation; FIC (Fuzzy Initial Conditions), that corresponds to a CDM simulation plus initial conditions from FDM, FIC+QP, for which we additionally activated the modified-gravity functions in \texttt{MG-PICOLA} to take into account the effects from the quantum potential $Q$; and EIC+QP (Extreme Initial Conditions), where extreme refers to the standard axion case and the reported excess of power in density perturbations~\cite{Cedeno2017,cedeo2020ultralight} with the modified-gravity functions activated. 

The main characteristics of our simulations are summarized in Table~\ref{tab:1}.
We performed a series of tests to assure the convergence of the numerical results, and from them, we selected the optimal parameters for the number of particles, number of mesh points, box size, the initial redshift of the simulations and time steps. For the later we use $800$ COLA steps given that the difference with a simulation of $1000$ steps is about $0.05 \%$. We use a fiducial boson mass of $m_a = 10^{-23} \, {\rm eV}/c^2$ for the FIC, FIC+QP and EIC+QP  simulations, in order to make sure the effects of the quantum terms for the range of wave numbers contained in our simulations boxes are well resolved, notice that the Nyquist scale is at $k_{\rm Nyquist}\simeq 214 h/{\rm Mpc}$. Additionally, our simulations were started at $z=30$ for the FIC and FIC+QP, and at $z=100$ for the EIC one, to assure the compatibility of the initial conditions with those of linear perturbations, which were obtained with an amended version of the Boltzmann code \texttt{CLASS} (v2.7)\cite{Urena-Lopez2016,Cedeno2017} (see also~\cite{Cookmeyer:2019rna} for a comparison of our formalism in solving the field equations with other approaches). 

\begin{table*}[htp!]
\caption{\label{tab:1} Parameter specification for the cosmological simulations described in the text. The Nyquist wave number for this setup is given by $k_{\rm Nyquist} \simeq 214 \, h/\mathrm{Mpc}$. The initial redshift for the CIC, FIC and FIC+QP is $z=30$, while for the EIC+QP is $z=100$. See the text for a discussion on the later case. }
\begin{ruledtabular}
\begin{tabular}{|c|c|c|c|c|c|}
 Type & Model  & Initial Conditions & $m_a$ (eV$/c^2$) & $N_{\rm part}$ & Box size (Mpc$/h$) \\
 \hline
CIC &$\Lambda$CDM & CDM   & -    &$ 1024^{3}$   &  15 \\
 FIC & $\Lambda$CDM & FDM &  $3\times 10^{-23} $ & $ 1024^{3}$   &  15\\
 FIC + QP & $\Lambda$FDM & FDM &  $ 3\times 10^{-23}$  &$ 1024^{3}$   &15\\
  EIC +QP   & $\Lambda$FDM & EFDM &  $3\times 10^{-23}$  & $ 1024^{3}$   &  15
\end{tabular}
\end{ruledtabular}
\end{table*}

\begin{figure*}[htp!] 
\centering
\includegraphics[width=1\textwidth]{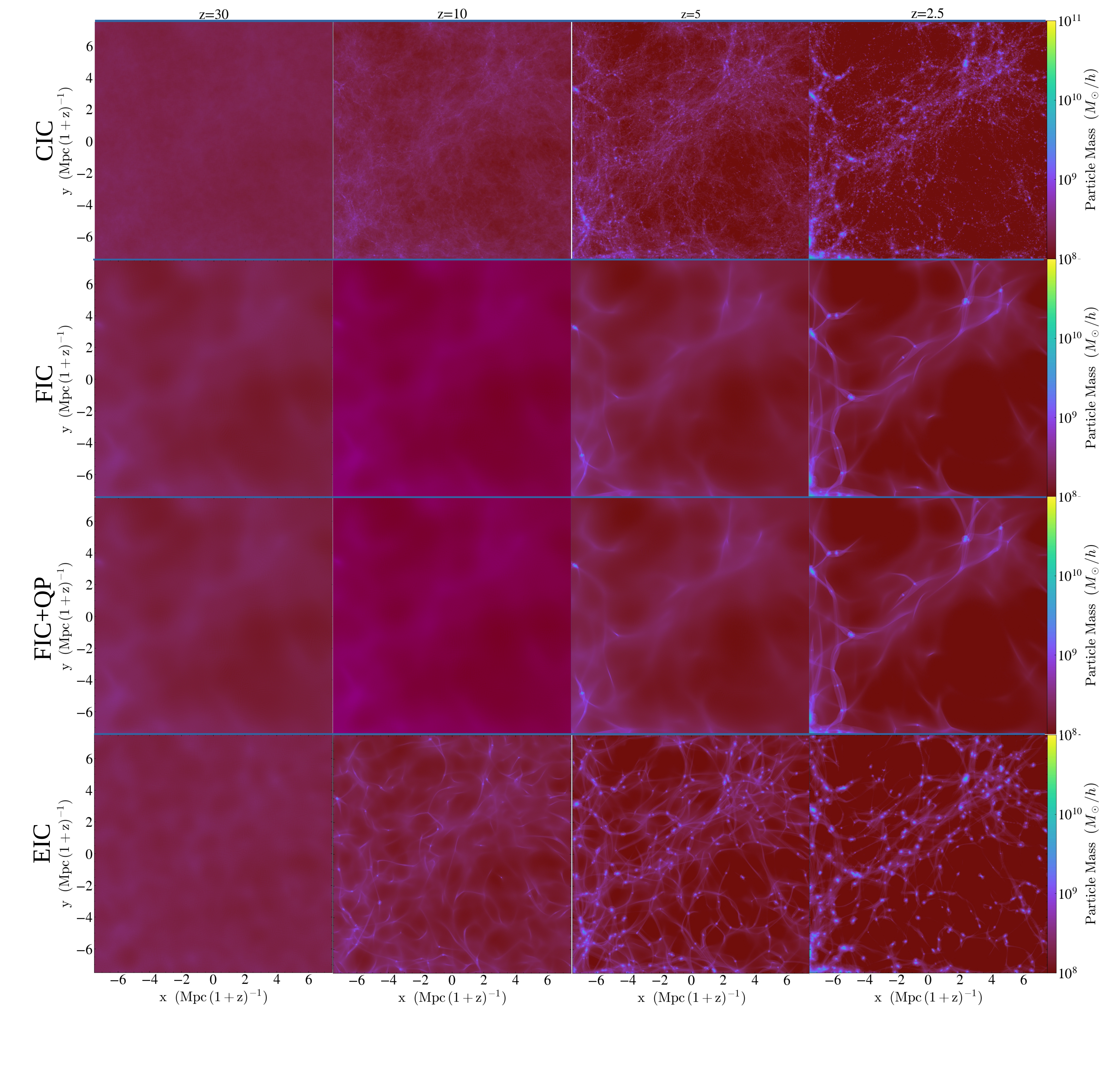}
\caption{\label{fig:grid_density}Density projection onto the plane $xy$ for the evolution of the cases FIC, FIC+QP, EIC, and CIC using the parameters shown in Table \ref{tab:1} and in comoving coordinates. See the text for more details.}
\end{figure*}

To begin with, in Fig.~\ref{fig:grid_density} we provide a visual comparison of the matter density projected onto the $xy$ plane, where the plots were generated with the \texttt{PYTHON} package \texttt{yt} \cite{2011ApJS..192....9T}. As reported in previous works~\cite{Zhang2017,Nori:2018pka}, it can be noticed that the simulations FIC and FIC+QP show less matter accumulation than the standard case CIC, an effect that must be attributed to the intrinsic properties of the boson system, Eq.~\eqref{eq:action}. However, the differences between the cases FIC and FIC+QP are difficult to spot by eye in the plots. In contrast, the simulation EIC shows more matter aggregation than the standard case CIC, which is in itself a manifestation of the power excess in the density perturbations already present in the initial conditions. 

\begin{figure}[tp!] 
\centering
\includegraphics[width=0.49\textwidth]{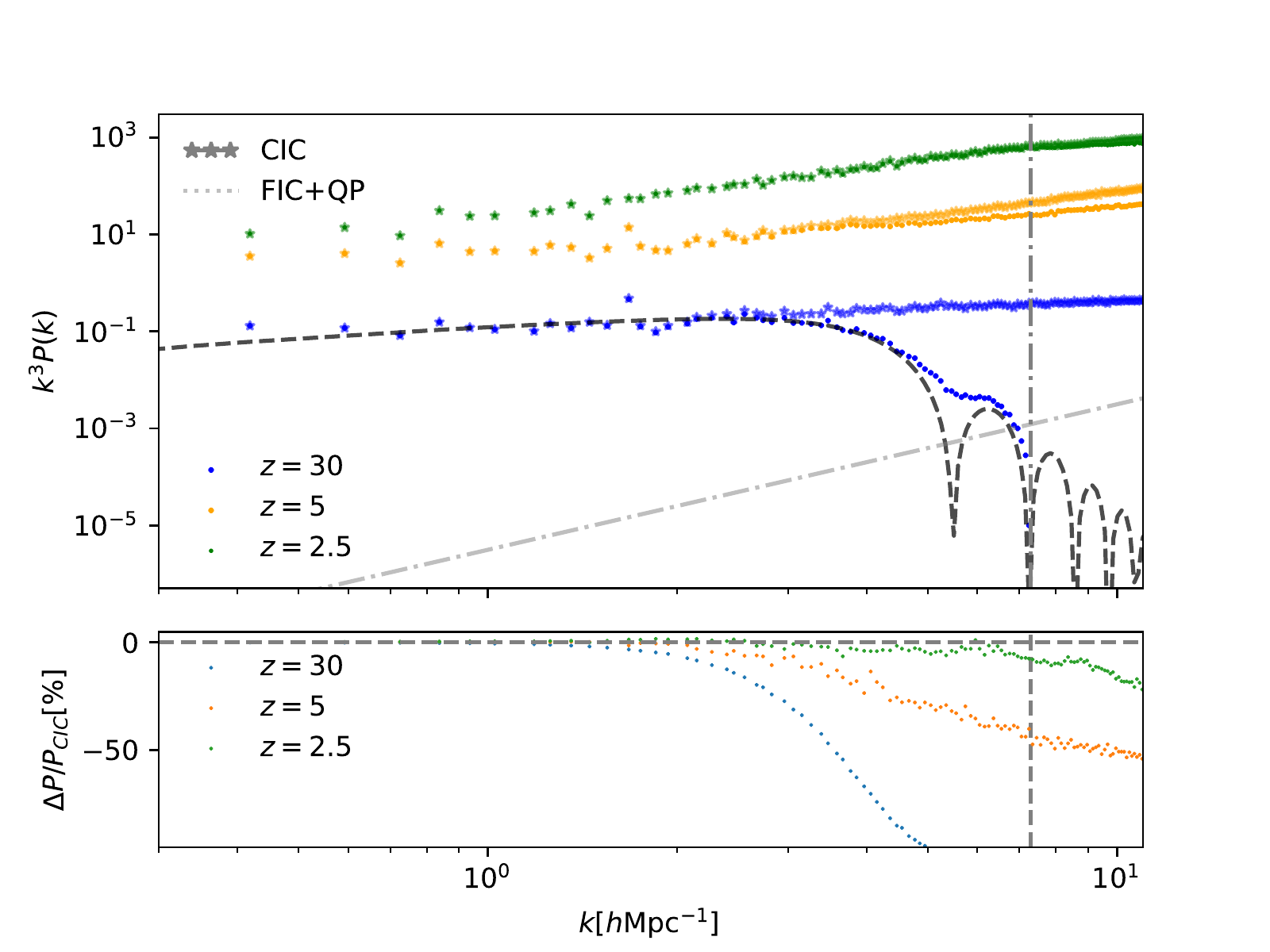}
\includegraphics[width=0.49\textwidth]{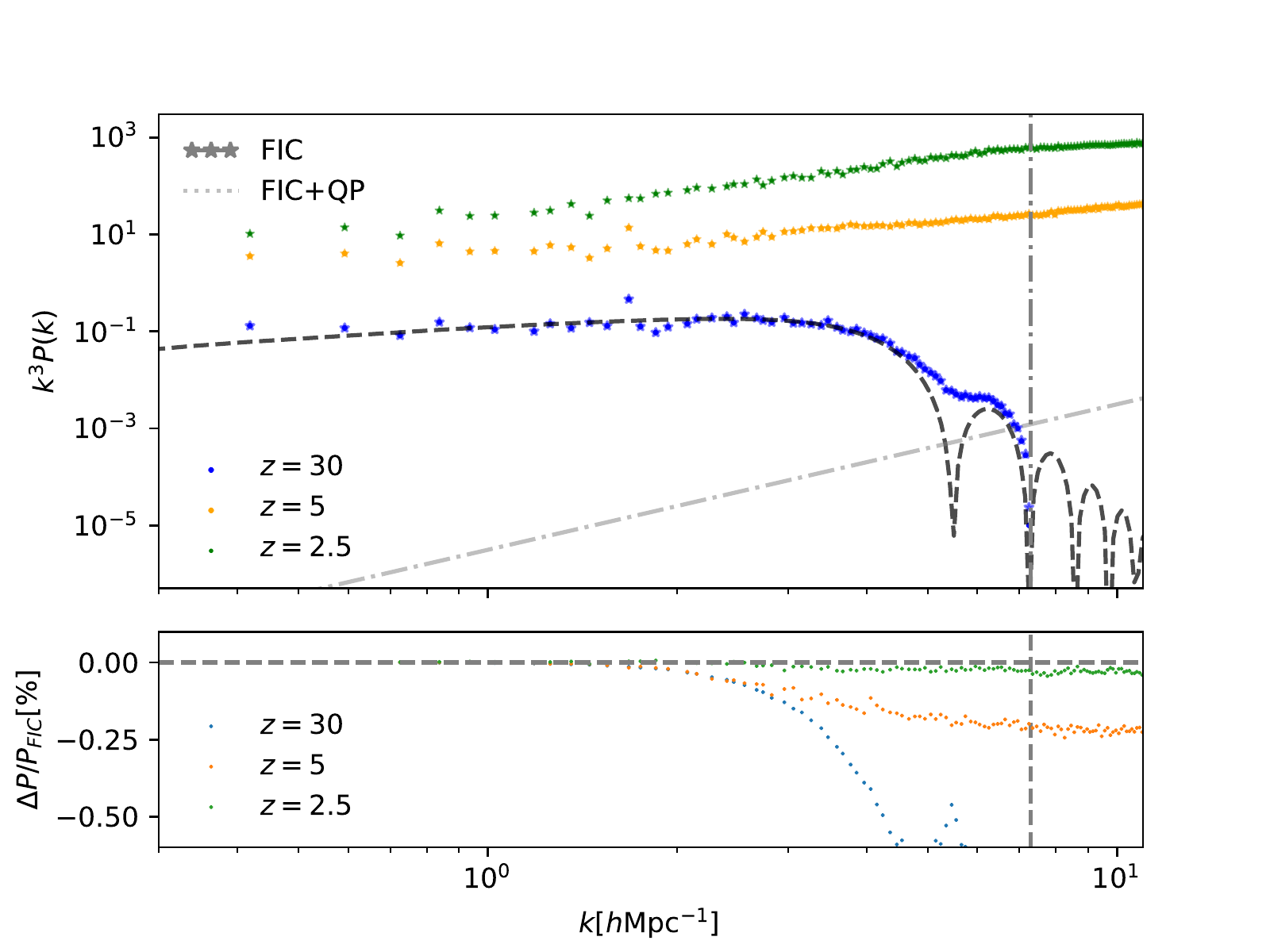}
\caption{\label{fig:back-a} 
Top: comparison of the measured particle MPS for the CIC (stars) and FIC+QP (dots) at different redshift, $z=30$ (blue), $z=5$ (yellow) and $z=2.5$ (green). The linear MPS from \texttt{CLASS}, used as initial condition, is shown for comparison (dashed black line). The lower sub--panel shows the relative difference between the cases considered, for the three redshifts:   $\Delta P/P_{\rm CIC} = (P_{\rm FIC+QP}-P_{\rm CIC})/P_{\rm CIC}$. Notice that the relative difference decreases for lower redshift.    We also show for reference the wave number for which the initial MPS is below the shot noise (vertical dot--dashed gray line) and the amplitude of the shot noise (sloping dot--dashed gray line), in order to delimit the $k$ range that could be influenced by shot noise. Bottom: the same as the top panel but the comparison here is with respect to the FIC simulation.}
\end{figure}

A more detailed analysis of the simulations was performed using the \texttt{PYTHON} package \texttt{NbodyKit}~\cite{nbody}, and the main quantity for the comparison between simulations is the MPS measured using the particles, which is shown in Fig.~\ref{fig:back-a} below for different redshift $z=30,5,2.5$ and for each simulation.

In the top panel of  Fig.~\ref{fig:back-a} we see a comparison between the simulations CIC, which is the standard CDM one, and FIC+QP, which considers both initial conditions and evolution of the FDM type. The initial MPS at $z=30$ for the latter shows the characteristic drop in power at small scales (for $k \gtrsim 2 \mathrm{Mpc}/h$) although a comparison with the linear MPS obtained directly from \texttt{CLASS} shows that the oscillations have been smoothed out by \texttt{MG-PICOLA}. Nonetheless, we see that the MPS of the FIC+QP simulation remains suppressed, at small scales, up to $z=5$, with respect to that of the CIC simulation, although the difference between the two is almost lost by $z=2.5$, being the relative difference of about $10\%$ at most.

We see in the FIC+QP simulations a transfer of power from large to small scales, an effect that has been widely reported in similar simulations of the FDM model \cite{Nori2018, Nori:2018pka, Li_2019}, although only for those that exploit the similarity with the hydrodynamic equations via the Madelung transformation. However, we consider the evolution of the MPS  to be trusted only up to $k \simeq 7.5 h \mathrm{Mpc}^{-1}$ (vertical dot--dashed line), which is  set by the scale at which the initial MPS is above the shot noise one (sloping dot--dashed line). Even though it is know that for particle-mesh like codes the shot noise affects less the evolution of models with suppression in the MPS~\cite{Angulo:2013sza}, we have preferred to be conservative. The sub--panel in this figure shows the relative difference between MPS measured in the CIC and FIC+QP, which we can notice diminishes with the redshift evolution. 

It should also be recalled that the Jeans length $k_{J0}$ is evolving, but for the parameters in our simulation its variation, from $z_i =30$ up to the present, is in the range (see for instance~\cite{Urena-Lopez2016,Marsh2016})
\begin{equation}
    13 \, h/ \mathrm{Mpc} \lesssim k_{J0} \lesssim 30 \, h/ \mathrm{Mpc} \, .
\end{equation}
This means that the SFDM density contrast, and in turn the MPS, is evolving like CDM for the range of wave numbers considered in Fig.~\ref{fig:back-a}, given that for them we obtain $\mu (a,k) \simeq 1$ for all practical purposes. 

As for the bottom panel of Fig.~\ref{fig:back-a}, we see a comparison between the FIC and FIC+QP simulations, to see the influence of the quantum potential term [see the last term on the rhs in Eq.~\eqref{eq:fluid-qbep-c}] in the evolution of the MPS. The initial MPS is the same for both simulations, and the relative difference is marginal afterwards, with a relative difference of about $0.25\%$ at most, with the MPS from the FIC+QP simulation consistently suppressed throughout the evolution, which we consider an effect of the quantum pressure (as suggested by the accompanying lower panel  which shows the fractional difference between the two cases). In overall, this comparison reinforces our statement above in that the inclusion of the quantum pressure gives marginal differences in the evolution of FDM density perturbations.\footnote{In Appendix~\ref{sec:comp_ax-gad} we show a comparison between simulations preformed with \texttt{MG-PICOLA} and the data shown in figure 9 of \cite{Nori2018}, which suggests that our results are in agreement with others reported in the literature.}

\begin{figure}[tp!] 
\centering
\includegraphics[width=0.49\textwidth]{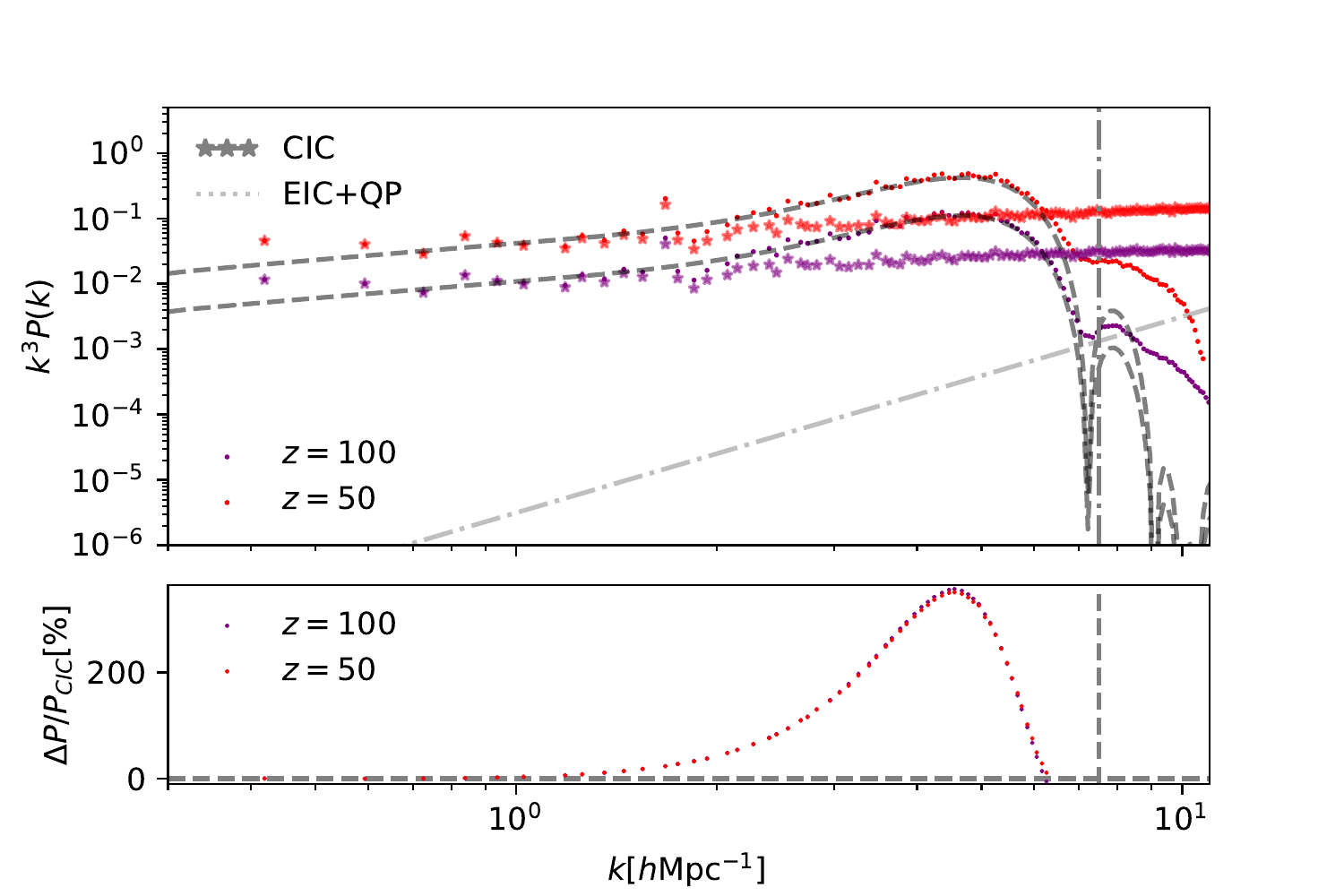}
\includegraphics[width=0.49\textwidth]{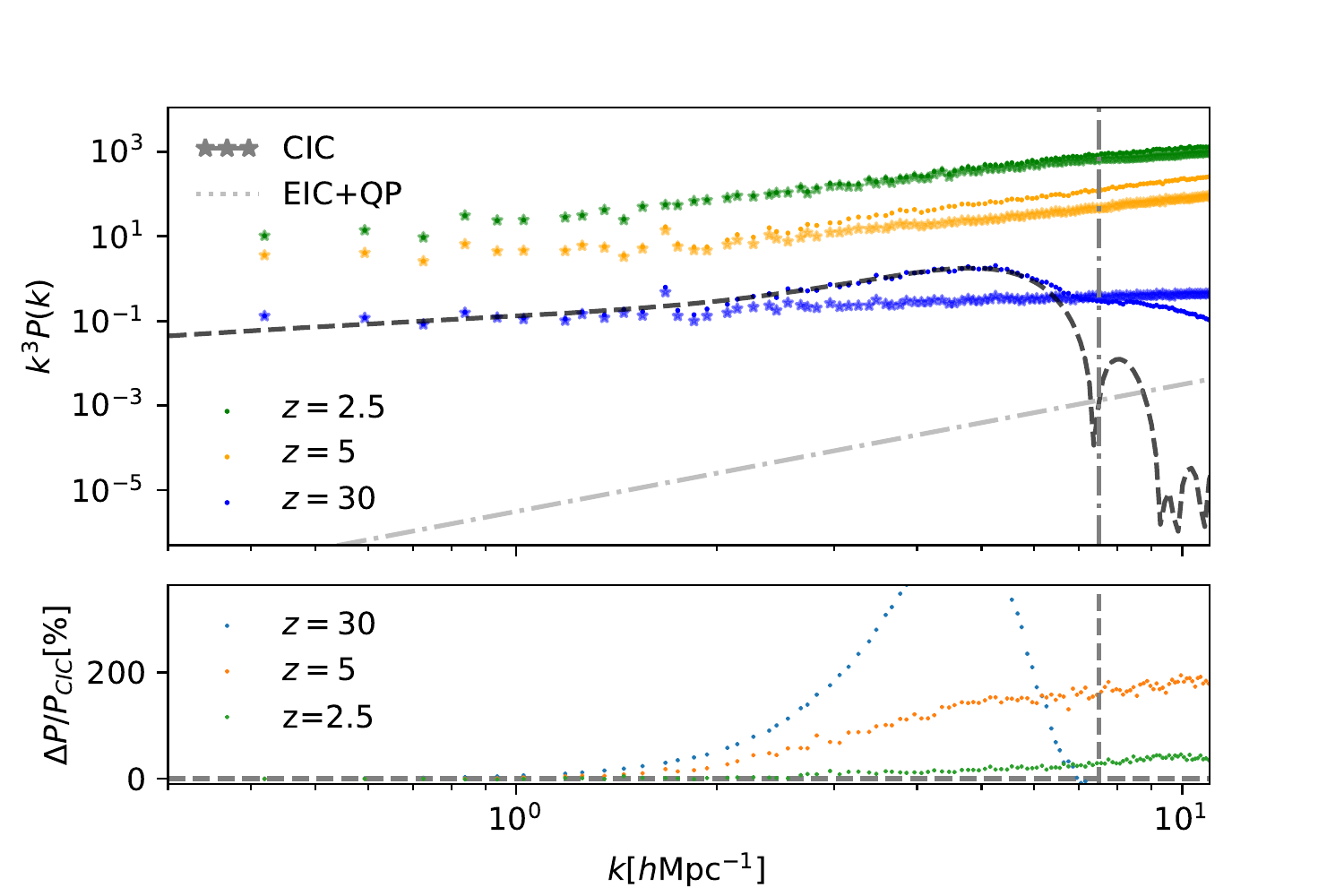}
\caption{\label{fig:back-b} 
(Top) Comparison of the measured particle MPS for the CIC (stars) and EIC+QP (dots) at different redshift, $z= 100$ (purple) and $z=50$ (red). See details of the simulation boxes in Table~\ref{tab:1}. The linear MPS from \texttt{CLASS}, used as initial condition, is shown for comparison (dashed black line). The lower sub--panel shows the relative difference between the cases considered, for the two redshifts:   $\Delta P/P_{\rm CIC} = (P_{\rm EIC+QP}-P_{\rm CIC})/P_{\rm CIC}$. Notice that the relative difference decreases for lower redshift. We also show for reference the wave number for which the initial MPS is below the shot noise (vertical dot--dashed gray line) and the amplitude of the shot noise (sloping dot--dashed gray line), in order to delimit the $k$ range that could be influenced by shot noise. (Bottom) The same as the top panel, but the comparison here is for the redshifts $z= 30$ (blue curve), $z= 5$ (yellow curve) and $z= 2.5$ (green curve). }
\end{figure}

Finally, in  Fig.~\ref{fig:back-b} we show the comparison between the CDM and EIC+QP simulations, the latter corresponding to the model with power enhancement at small scales. Here, the acronym EIC+QP means extreme initial conditions plus the quantum force, where extreme refers to the standard axion case and the reported excess of power in density perturbations reported in~\cite{Zhang2017,Cedeno2017,cedeo2020ultralight}. As explained in~\cite{Cedeno2017,cedeo2020ultralight}, the axion case can be approximated by the action ~\eqref{eq:action} but with a negative self-interaction parameter, $\tilde{\lambda} < 0$.\footnote{This also means that $k^2_{J1} \to - k^2_{J1}$ in Eq.~\eqref{eq:qbep-mgcola}, and this opens the possibility that $\mu(a,k,\tilde{\lambda}) \gtrsim 1$ at early times in the evolution of $\delta_k$, see Eq.~\eqref{eq:qbep-mgcola-a}. This would be the underlying mechanism for the excess of power with respect to CDM in the case of axion density perturbations. However, see~\cite{Zhang2017,Cedeno2017,cedeo2020ultralight} for a wider explanation.} Actually, the self-interaction term in the axion potential can be written as $\tilde{\lambda} = -(\tilde{m}^2_a/4!\tilde{f}^2_a)$, where $\tilde{f}_a$ is the so-called axion decay constant.\footnote{ Notice that this allows us to write $k_{J1} = 6 a \, \tilde{m}_a (8 \pi G \tilde{f}^2_a/3c^4)^{1/2}$, which shows that the second Jeans wave number is basically set up by the Compton length of the boson (axion) particle.}

We can also consider such a case given that the excess of power must be imprinted on the MPS around the time the scalar field behaves as a pressureless matter, which happens before the time of radiation-matter equality if $m_a > 10^{-25} \, \mathrm{eV} /c^2$~\cite{Urena-Lopez2016}. Thus, the only effect that is still present at late times is the suppression of the MPS at small scales due to the quantum potential term in Eq.~\eqref{eq:fluid-qbep-c}. In other words, EIC+QP means that the initial MPS has an excess of power with respect to CDM and that density perturbations are evolved by the same dynamics as the FDM model.

The EIC+QP case shown in Fig.~\ref{fig:back-b} considers that the decay constant of the axion field $\tilde{f}_a$ is such that $\sqrt{8 \pi G} \tilde{f}_a/c^2 \simeq 1.94 \times 10^{-3}$, and the initial MPS was calculated as explained in~\cite{Cedeno2017,cedeo2020ultralight}. We can see that at $z=100$ there is already an excess of power with respect to CDM at around $k \simeq 7 \, h \mathrm{Mpc}^{-1}$, and also a typical suppression of power for larger wave numbers (smaller scales).

As mentioned before for the previous cases, the initial MPS calculated by \texttt{MG-PICOLA} appears to be smoothed out with respect to that obtained from \texttt{CLASS}, but otherwise they agree very well. However, we can see that by $z=30$ there appears a noticeable difference in the MPS calculated from the two codes: the one from  \texttt{MG-PICOLA} shows a less pronounced suppression of power at small scales. This result is in stark contrast to the cases in Fig.~\ref{fig:back-a}, for which the agreement between the two MPS is very good at $z=30$. Such result can be explained in terms of the transfer of power from large to small scales, due to the presence of the power excess in the initial MPS, but such transfer seems to happen from a much earlier time.

A similar effect was studied before in ~\cite{Ma_2007} for artificially induced peaks in the initial MPS: the nonlinear MPS obtained from $N$-body simulations shows that power from large scales is transferred evenly to small scales. In our case, which in addition to a peak has a suppression of power at small scales, such effect means that the nonlinear MPS in the EIC+QP case should show an excess of power at small scales as compared to the standard FIC case.

\section{Conclusions and discussion \label{sec:conclusions}}
There are some limitations in our approach to SFDM simulations. The first one is that in the calculation of the scale-dependent corrections to the evolution of the first and second order growth factors, we had to rely on perturbative expansions in terms of $\delta$ for the effective quantum potential. Such expansion is formally valid for small amplitudes $|\delta| \lesssim 1$, and then is not reliable for the highly nonlinear regime of density perturbations. This is a limitation of the SFDM model only, as for Modified Gravity models the scale-dependent corrections are directly proportional to the density contrast $\delta$ by definition and then all calculations are applicable to both the linear and nonlinear regimes.

Nonetheless, our results are to be trusted up to second order in $\delta$, specially because the second order corrections were canceled by the particular approximation of orthogonal triangle configurations in Fourier space. This is a favorable coincidence that helped us to have more reliable results at least at the level of 2LPT (see also Appendix~\ref{sec:comp_ax-gad}). 

One general result in our simulations is the transfer of power from large to small scales, as seen in the calculated MPS, which means that the characteristic suppression of power is no longer as acute as in the linear evolution, although we were able to quantify a visible difference with respect to CDM even at small redshifts. It should be noted that our numerical simulations do not show any major difference in the aforementioned transfer of power in the cases we called the CDM and FDM evolution (FIC and FIC+QP simulations, respectively), and then the final profile of the MPS is the same for all practical purposes up to redshift $z=2.5$. 

We also presented the results corresponding to the extreme axion case (EIC+QP), with its characteristic excess of power in the MPS. Here our simulations indicate also a convergence of the MPS toward the CDM one by means of transfer of power from large to small scales, although the MPS remains overpowered as compared to CDM.

It is important to mention that although it is well known that for models with suppression in the MPS the effect of the shot noise can be important, it is also known that particle-mesh like techniques are less affected \cite{Angulo:2013sza}. Nevertheless, in order to use this type of simulations for larger axion mass values, which are of higher interest given current constraints, it would require us to make a deeper study of the effect of the noise that we expect to present elsewhere.  

\begin{acknowledgments}
We thank Jorge Cervantes-Cota and Alejandro Avil\'es for useful conversations and clarifications on the scale-dependent COLA method, Octavio Valenzuela for the discussion of the results from the simulations perspective, and Oleg Burgue\~no for technical support with the COUGHs server where the simulations for this work were performed. We are grateful to Matteo Nori and Marco Baldi for kindly providing some of the data from Ref.~\cite{Nori:2018pka} to generate Fig.~\ref{fig:com-axz49}. SGM-G thanks CONACYT for financial support. A.X.G.M. acknowledges support from C\'atedras CONACYT. This work was partially supported by Programa para el Desarrollo Profesional Docente; Direcci\'on de Apoyo a la Investigaci\'on y al Posgrado, Universidad de Guanajuato, research Grants No. 036/2020, 099/2020; CONACyT M\'exico under Grants No. A1-S-17899, No. 286897, No. 297771, No. 304001; and the Instituto Avanzado de Cosmolog\'ia Collaboration.
\end{acknowledgments}

\appendix

\section{Comparison with AX-GADGET}
\label{sec:comp_ax-gad}
\begin{figure*}[tp!] 
\centering
\includegraphics[width=0.49\textwidth]{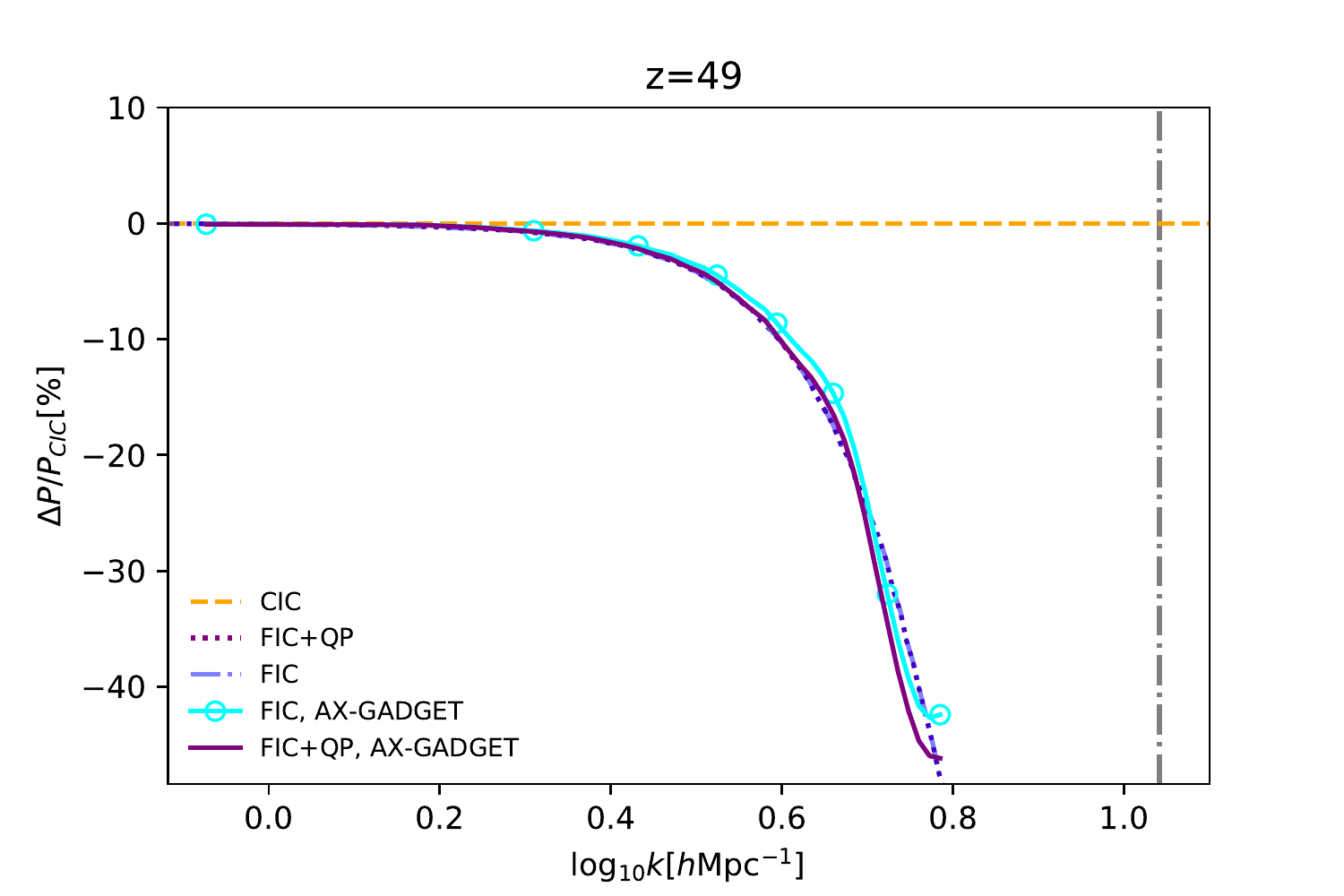}
\includegraphics[width=0.49\textwidth]{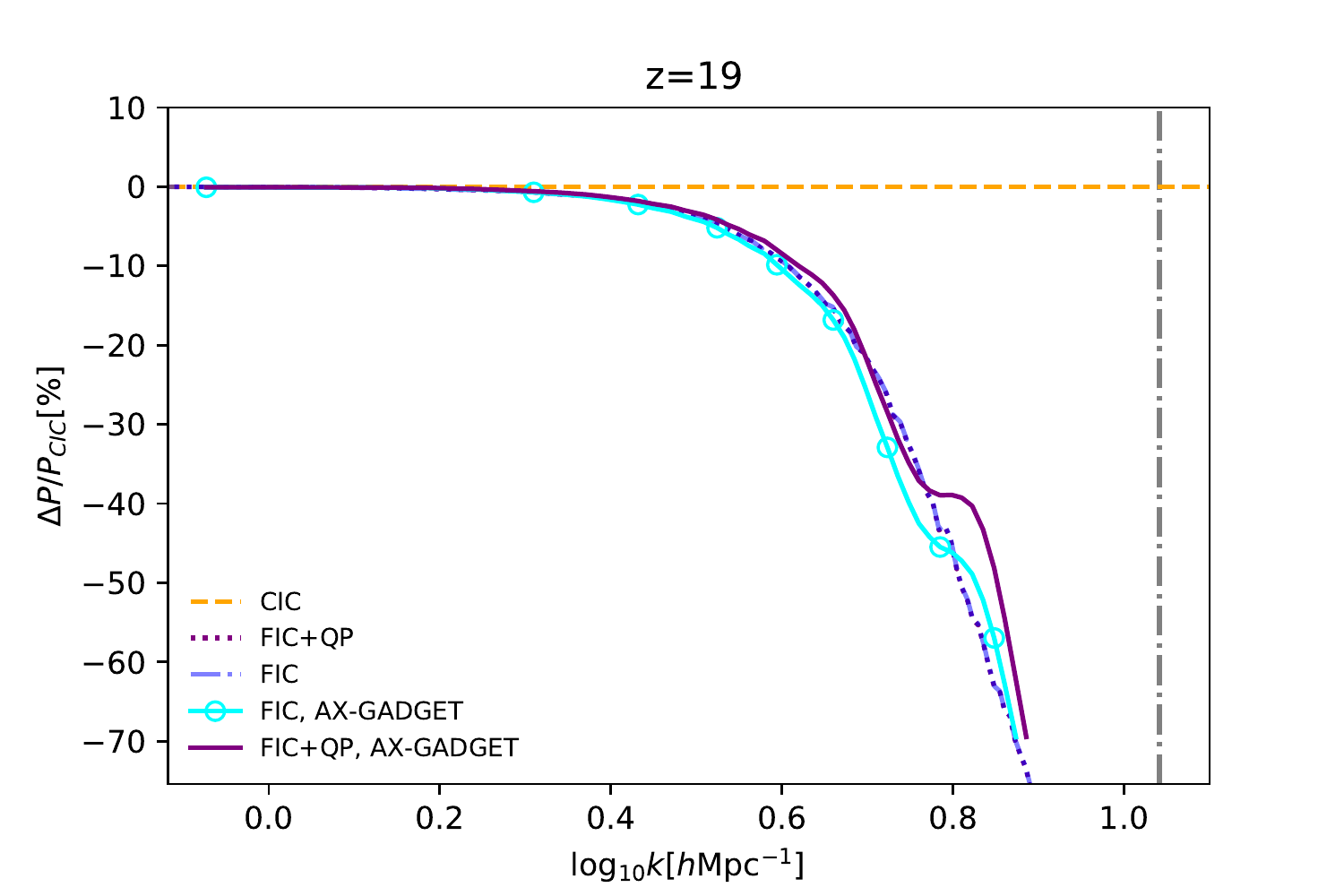}
\includegraphics[width=0.49\textwidth]{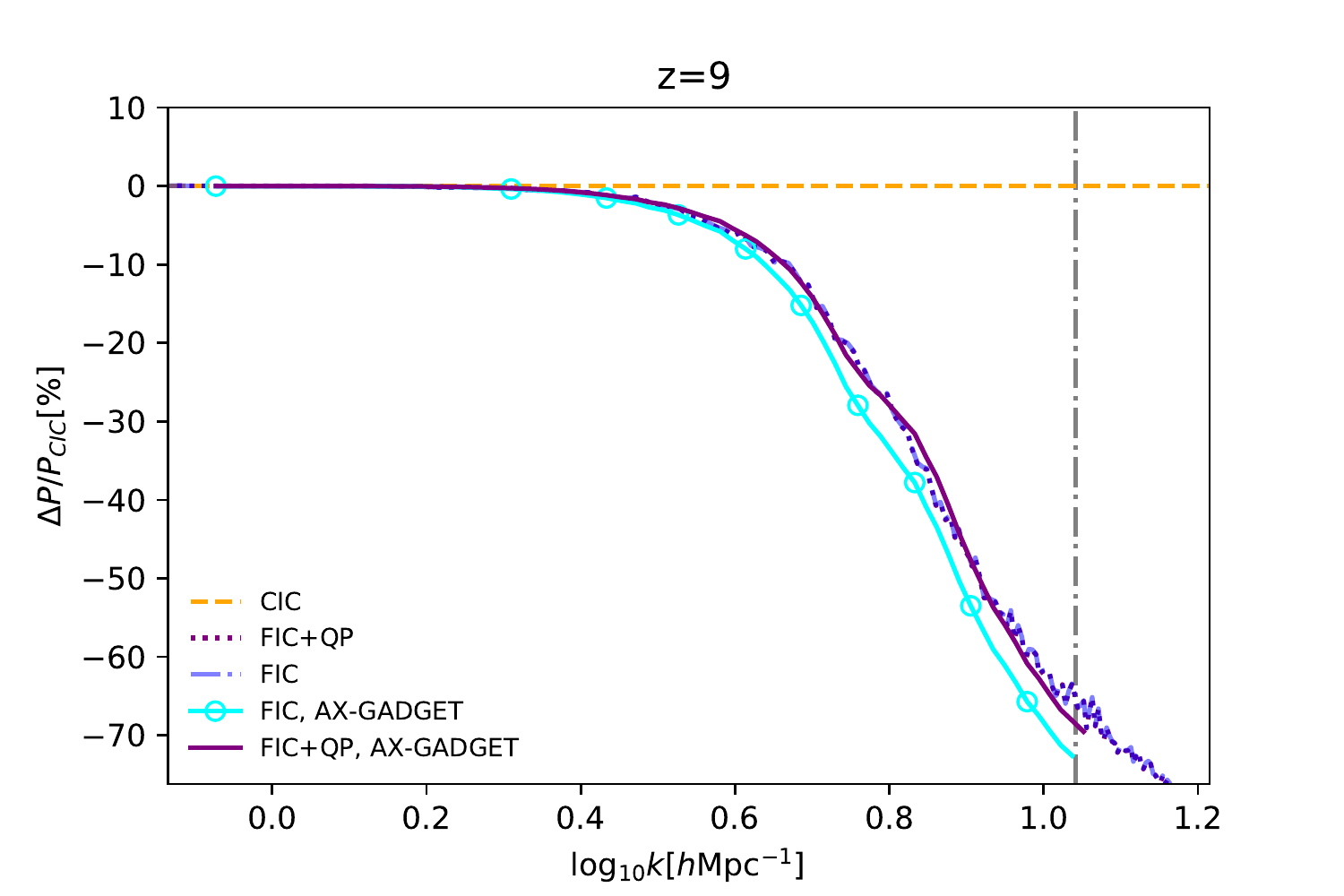}
\includegraphics[width=0.49\textwidth]{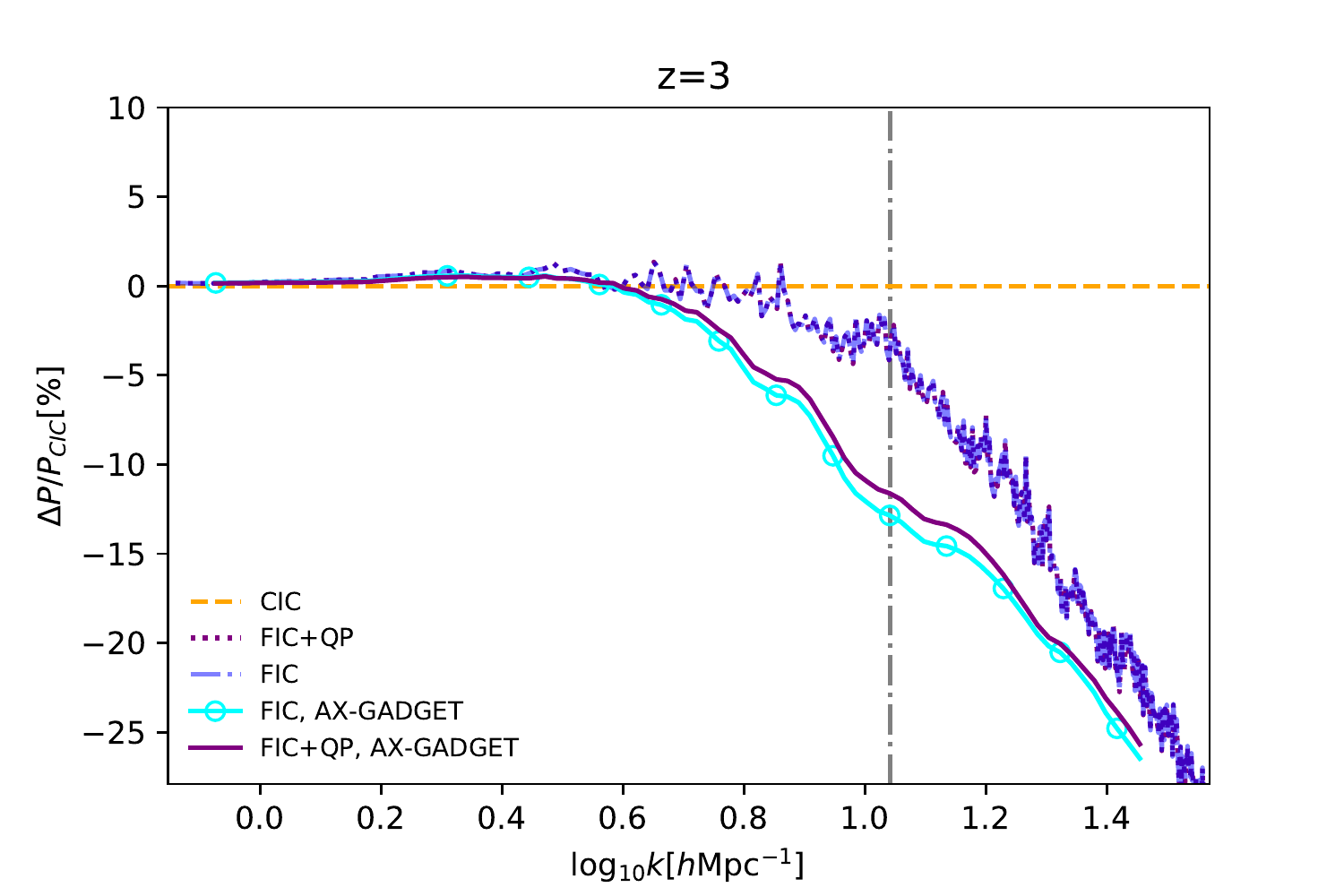}
\caption{\label{fig:com-axz49} Comparison between the results of the simulations shown in Fig.~9 of Ref.~\cite{Nori2018,private} (\texttt{AX-GADGET}) and simulations with \texttt{MG-PICOLA}. The parameters in the simulations are $m_a= 10^{-22} \mathrm{eV}/c^2$, a box size of $15 \, \mathrm{Mpc}/h$ and $512^{3}$ particles. As explained in Sec.~\ref{sec:cosmo-sims}, the results from the cases FIC and FIC+QP are almost indistinguishable in the plots. For reference we also show the wave number for which the initial MPS is below the shot noise (vertical dot-dashed gray line). See the text for more details.   
}
\end{figure*}

We show in Fig.~\ref{fig:com-axz49} the relative differences in the MPS for the cases FIC and FIC+QP in Fig.~9 of Ref.~\cite{Nori2018} obtained from the code \texttt{AX-GADGET}~\cite{private}, in comparison with our results using the amended code \texttt{MG-PICOLA} as explained in the main text. The relative difference is calculated with respect to the standard case CIC. 

The parameters in our simulations in Fig.~\ref{fig:com-axz49}, corresponding to number of particles, box size, boson mass and initial redshift $z=50$, were chosen to match those of the simulations in~\cite{Nori2018}. It can be seen that the agreement is very good in the cases FIC and FIC+QP, except at the lowest redshift $z=3$ for which the results from \texttt{AX-GADGET} appear a bit more suppressed, which may suggest an extra influence of the quantum potential at low redshifts.

As mentioned in Sec.~\ref{sec:conclusions}, this may be part of the limitations in our approach, and probably a third order approximation LPT is necessary to reach more accuracy. However, this study is out of the scope of this work and we expect to report it elsewhere.

\bibliography{extremesfdm}

\end{document}